\documentclass[prd,showpacs,showkeys,floatfix,nofootinbib,
               eqsecnum,
               preprint,12pt,tightenlines,fleqn]{revtex4} 

\usepackage{amsmath,amssymb,revsymb,graphicx,dcolumn}

\newcommand  {\version}{v6}  

\begin{document}

\hfill (\version; \today)\newline\vspace*{1\baselineskip}
\title[$\hbar$ as parameter of Minkowski metric in effective theory]
        {$\hbar$ as parameter of Minkowski metric  in effective theory
        \vspace*{.5\baselineskip}}
\author{G.E. Volovik
\\
Low Temperature Laboratory, Helsinki University of Technology, 
P.O.Box 5100, FIN-02015, HUT, Finland
\\
 Landau Institute for Theoretical Physics RAS, Kosygina 2,
119334 Moscow, Russia
 \vspace*{.5\baselineskip}}
\begin{abstract}
\vspace*{1.0\baselineskip}\noindent
With the proper choice of the dimensionality of the metric components matter field 
variables,
the action for all fields becomes dimensionless. Such quantities as
the vacuum speed of light $c$,
the Planck constant $\hbar$, the electric charge $e$,
the particle mass $m$, the Newton constant $G$ never enter equations written
in the covariant form, i.e., via the metric $g^{\mu\nu}$.
The speed of light  $c$ and the
Planck constant $\hbar$ are parameters of
 a particular two-parametric family of solutions of general relativity
 equations describing the flat isotropic Minkowski vacuum in effective theory emerging at low energy:
 $g^{\mu\nu}_\mathrm{Minkowski}=
 \mathrm{diag}(-\hbar^2, (\hbar c)^2,  (\hbar c)^2, (\hbar c)^2)$.
 They parametrize the equilibrium quantum vacuum state.
 The physical quantities which enter the covariant equations
are dimensionless quantities and dimensionful quantities of dimension of rest energy $M$
or its power.
Dimensionless quantities include  the running coupling `constants'  $\alpha_i$; 
the  geometric $\theta$-parameters which enter topological terms in action; 
and geometric charges coming from the group theory, such as
angular momentum quantum number $j$, weak charge, electric charge $q$, hypercharge,  baryonic
and leptonic charges, number of atoms $N$, etc.
Dimensionful parameters have dimensions of the rest energy and its powers. They include the rest energies of particles $M_n$ (or/and mass matrices); the  gravitational coupling $K$  with dimension of $M^{2}$;  string tension in QCD $K_\text{QCD}$ with dimension $M^2$; cosmological constant with dimension $M^4$; etc. 
In effective theory, the interval  $s$  has the dimension of $1/M$; it characterizes
the dynamics of particles in the quantum vacuum rather than
the  geometry of space-time. The action is dimensionless reflecting 
 equivalence between an action
and the phase of a wave function in quantum mechanics.
We discuss the effective action, and the measured physical quantities resulting from the action, including parameters of metrology triangle which enter the Josepson effect, quantum Hall effect, and quantum pumping.

\end{abstract}

\maketitle

\section{Introduction}

The system of units is based on the theoretical understanding of physical laws. The traditional approach to the system of units is based on the two great physical theories of the twentieth century: special relativity and quantum mechanics, which suggest to  fix the speed of light  $c$ to connect
space and time units,  and the Planck constant $\hbar$ to connect mass and time units \cite{Borde2005}. In theoretical physics, these quantities are often considered as fundamental constants and are used as units: in these units,  $c=\hbar=1$
\cite{Uzan2003}.

However, another great theory --  general relativity (GR) -- undermines this approach. 

\subsection{Speed of light}

In mechanics (Galilean, relativistic or any other) the energy of a freely moving body in the limit of small momentum is expanded in terms of momentum:
\begin{equation}
 E({\bf p})=M +\frac{1}{2}K^{ik}p_i p_k +K^{ikmn}p_i p_k p_m p_n +...
\label{eq:general_energy}
\end{equation}
In the Galilean invariant mechanics, equation \eqref{eq:general_energy} contains only the first two terms, the internal or the rest energy $M$ and kinetic energy with the
isotropic mass tensor $K^{ik}=m^{-1}\delta^{ik}$:
\begin{equation}
 E({\bf p})=M +\frac{{\bf p}^2}{2m} \,.
\label{eq:Classical_energy}
\end{equation}
It contains only two parameters of the body: the rest energy $M$ and the 
inertial mass $m$ in the kinetic energy.   The Lorentz invariant theory -- special relativity -- connects two parameters of the body, inertial mass $m$ and  the rest energy $M$,  via the speed of light,
\begin{equation}
 m=\frac {M}{c^2}\,,
\label{eq:Energy_mass_relation}
\end{equation}
and equation \eqref{eq:general_energy} is transformed to equation with
the parameter $M$ and `fundamental constant' $c$:
\begin{equation}
E^2-c^2p^2-M^2=0\,.
\label{eq:Special_relativity}
\end{equation}

GR effectively removes the `fundamental constant' from the equation
\eqref{eq:Special_relativity}, it transforms this equation to
\begin{equation}
 g^{\mu\nu}p_\mu p_\nu + M^2=0 ~~,~~p_\mu=(-E,p_i) \,.
\label{eq:GR_energy}
\end{equation}
Equation \eqref{eq:GR_energy} contains only the parameter of a body, the rest energy $M$. The speed of light $c$ does not enter explicitly: it becomes the part of the metric. Because of that, $c$ never enters explicitly any equation, which
is written in the covariant form, i.e. when the equation is expressed in terms of metric field (see e.g. \cite{Volovik2003,Volovik2002}). It may  enter  only the  solutions of equations, in particular the solutions which have Minkowski space-time asymptotically.

In GR, the relation between space and time is not universal but depends on
the metric field $g_{\mu\nu}({\bf r},t)$. The speed of light $c$, by definition of speed
as an infinitesimal proper length divided by an infinitesimal proper time, is determined
 in the limit of vanishing distance between the point objects. But
zero distance limit is mathematical construction which does not reflect the real physical world.  
The physically measured speed of propagation of light between two distant objects is coordinate
dependent and thus depends on the trajectory of propagation.
This implies that the parameter $c$ may make sense only in the Minkowski space-time,
when gravity is fully absent:
\begin{equation}
 g_{\mu\nu}^\mathrm{Minkowski}\left(c\right)=
 \mathrm{diag}\left(-1,  \frac{1}{c^2},    \frac{1}{c^2}, \frac{1}{c^2}\right) \,.
\label{eq:Isotropic_Minkowski}
\end{equation}
In fundamental GR the limiting speed $c$ is fundamental, which allows us to put $c=1$.  

In effective theory, the latter is problematic. For example, if the underlying 
microscopic system  is anisotropic, the  limiting speed  of the low-energy excitations
depends on the direction of propagation, and effective Minkowski space-time contains 3 parameters:
\begin{equation}
 g_{\mu\nu}^\mathrm{Minkowski}\left(c_1,c_2,c_3\right)=
 \mathrm{diag}\left(-1, \frac{1}{c_1^2},   \frac{1}{c_2^2},  \frac{1}{c_3^2}\right) \,.
\label{eq:Anisotropic_Minkowski}
\end{equation}
 Such anisotropy of the physical speed of light will be revealed only at high energy.  For the internal  low-energy observers,  the world obeys Lorentz invariance and equivalence principle of GR. The measured limiting speed  is coordinate independent, isotropic and universal for all species 
(at least in the Fermi-point scenario of emergent gravity \cite{Volovik2003}), if  observers
do not use ``xylophones, yachts and zebras
to measure intervals along the $x$, $y$ and $z$ axes''  \cite{Trialogue},  but use the same xylophone in all directions.  Since xylophones (rods and clocks) are made of anisotropic quasipaticles, the  rescaling of measured time and distances automatically occurs due to the physical Lorentz--Fitzgerald contraction
of length of   rods and
the physical Lorentz slowing down of   clocks. This  leads to apparent isotropy and universality in the low-energy limit, while for external high-energy observer the speed of light depends on coordinates and energy and is different for different species. 

Thus in both cases, fundamental and emergent, the limiting speed $c$ drops out of any equation written in the covariant form. For fundamental GR this is trivial since one may put  $c=1$. In effective GR even such notion as the fundamental speed is simply absent and its introduction is artificial. Let us now show that the same occurs with another fundamental constant, the  $\hbar$.

\subsection{Planck constant $\hbar$}

Planck constant relates the frequency of emitted photon with the energy levels of atom:
\begin{equation}
 M_m - M_n=\hbar \omega_{mn} \,.
\label{eq:Emission}
\end{equation}
For an extended body the relation between the invariant mass $M$ and the non-covariant frequency 
is also coordinate dependent due to gravitational red shift. This means
that the parameter $\hbar$ can be measured only in the ideal limit of a point object. The string theory deals with the extended objects --  fundamental strings, and in effective GR the size  of an object is also limited at least by Planck length.
This suggests that the quantity $\hbar$ is also the characteristic of  Minkowski space-time and can be absorbed by metric  as $c$.

Such absorption occurs, if the quantity $\hbar$, which traditionally is the prefactor in the quantum mechanical operator of momentum $p_\mu=-i\hbar\nabla_\mu$, is 
moved from $p_\mu$ to the metric $g_{\mu\nu}$ in Eq. \eqref{eq:GR_energy}. As a result,  the isotropic Minkowski metric is characterized by two parameters, $\hbar$ and $c$, with $\hbar$ being the conformal factor of the Minkowski space-time:
\begin{equation}
 g_{\mu\nu}^\mathrm{Minkowski}\left(c,\hbar\right)=
 \hbar^{-2}\mathrm{diag}\left(-1,  c^{-2},  c^{-2},   c^{-2}\right)~~,~~ g^{\mu\nu}_\mathrm{Minkowski}\left(c,\hbar\right)=
 \hbar^2\mathrm{diag}\left(-1,  c^2,  c^2,  c^2\right)
 \,.
\label{eq:Minkowski_metric}
\end{equation}
Now when  $\hbar$ is absorbed by the metric,
it also does not enter explicitly any covariant equation. 
In particular,  equation \eqref{eq:Emission} becomes
\begin{equation}
 M_m - M_n=\frac{\omega_{mn}}{\sqrt{g_{00}}} \,,
\label{eq:Emission_covariant}
\end{equation}
 i.e. in GR the quantum mechanic equation \eqref{eq:Emission} is hidden 
 in the equation \eqref{eq:Emission_covariant} describing the red shift.

With the choice of the dimensionality of the metric components in Eq. \eqref{eq:Minkowski_metric}, 
 the  action becomes dimensionless,  $[S]=1$. An example is  the
 the classical action for a freely moving massive particle:
\begin{equation}
S_M=M\int ds~~,~~ds^2=g_{\mu\nu}dx^\mu dx^\nu \,.
\label{eq:Particle_Action}
\end{equation}
This action is dimensionless. A dimensionless action leads to a natural
formulation of quantum mechanics
in terms of Feynman path integral with the integrand  $e^{iS}$.
For a single particle, $S$ is the phase of a semiclassical
wave function, while equation \eqref{eq:GR_energy} transforms to
\begin{equation}
 g^{\mu\nu}k_\mu k_\nu + M^2=0 ~~,~~k_\mu=(-\omega, k_i) \,.
\label{eq:GR_frequency}
\end{equation}

If gravity is fundamental and $\hbar$ is fundamental constant, the proposal to eliminate $\hbar$ from equations by hiding it into the metric \eqref{eq:Minkowski_metric} by  scale transformation  would be a formal mathematical 
trick only.  Nothing prevents to use the scale factor  larger or smaller than $\hbar$, if it is accompanied by the rescaling of mass $M$.  
If all the equations are covariant, one may safely put $c=\hbar=1$.

However,  if gravity is effective and general covariance emerges together with gravity  only in the low-energy corner of  the underlying microscopic physics of quantum vacuum, then all 10 components of the metric tensor become physical including the parameters $c$ and $\hbar$, entering the Minkowski metric of the equilibrium vacuum state. They are not universal and together with the other parameters of the metric may explicitly enter  the higher order terms in the action, for which the covariance as emergent phenomenon is not valid.  As a results the ratio  $(M_m-M_n)/\omega_{mn}$, which is universal in the low-energy range, will depend on frequency at high energy. 
This makes impossible setting $\hbar=1$, because this quantity depends on the vacuum state and in principle is time and coordinate dependent.

An example is provided by the $q$-theory \cite{KlinkhamerVolovik2008b} of quantum vacuum. 
According to the $q$-theory, our quantum vacuum is a self-sustained medium. It has an  
equilibrium state in which the parameter $q$ is self-adjusted to its equilbirium value $q_0$. It was recently demonstrated that such a self-adjusted equilibrium state may serve as an attractor, i.e. the non-equilibrium vacua relax towards the flat space-time with this value of $q$ \cite{KlinkhamerVolovik2009c}. 
In principle it is possible that in emergent theory the Universe relaxes to the distinguished Minkowski space-time with fixed values of parameters  $c(q_0)$ and $\hbar(q_0)$.  Such an equilibrium vacuum could be the final state  which the system approaches in our part of the Universe.

In this paper we restrict ourselves with the covariant equations, i.e. 
we consider the lowest order terms in the effective theory and neglect the higher-order terms, which
violate the emergent low-energy symmetry. We demonstrate
that the choice \eqref{eq:Minkowski_metric}
for  Minkowski metric allows us to remove $c$ and $\hbar$ from all the covariant equations without
setting $c=\hbar=1$, i.e. without assumption that these quantity are universal and coordinate independent. On the contrary, when $c$ and $\hbar$ are inside the metric, the symmetry arguments (gauge invariance, general covariance, etc.) do not  prohibit the space-time dependence of  $c$ and $\hbar$.   This is possibly the necessary intermediate step towards the ``quantum gravity'', the underlying microscopic theory in which  $c$ and $\hbar$ may and should depend on space and time.

\section{Effective action}

Let us first discuss the effective action for the gauge fields and gravity as it
appears, say, in the Fermi-point scenario \cite{Volovik2003}, and rewrite it in such a way
that $\hbar$ is hidden in the metric field. The main lesson is that in the low-energy corner, i.e. in the region of applicability of effective theory, the parameter $\hbar$ never enters explicitly.

\subsection{Electromagnetic action}

We choose the dimensions of the vector potential $A_\mu$
and field strength $F_{\mu\nu}$ as they
naturally follow from the
geometric origin of the gauge field. Since $A_\mu$  arises as a result of localization of the dimensionless $U(1)$ field, $\nabla_\mu \phi \rightarrow A_\mu$, one has:
 \begin{equation}
[A_0]=[t]^{-1} ~~,~~[A_i]=[l]^{-1} ~~,~~[F_{ik}]=[l]^{-2}~~,
~~[F_{i0}]=[t]^{-1} [l]^{-1} \,.
\label{eq:em_dim}
\end{equation}

The interaction of a classical particle 
with electromagnetic field is
\begin{equation}
S_\text{int}=q \int dx^\mu A_\mu  \,.
\label{eq:interaction}
\end{equation}
Here $q$ is the geometric charge of a particle corresponding to the $U(1)$ gauge group, with $q_e=-1$ for electron; $q_u=2/3$ for up quark;  $q_d=-1/3$ for down quark, etc. In quantum mechanics, equation \eqref{eq:interaction} corresponds to the covariant derivative $D_\mu =\nabla_\mu - iq A_\mu$.
The motion equation of a massive particle with electric charge $q$ in electromagnetic field follows  from \eqref{eq:Particle_Action} and \eqref{eq:interaction}:
\begin{equation}
\frac{du^\mu}{ds}+\Gamma^\mu_{\lambda\sigma} u^\lambda u^\sigma  =
\frac{q}{M} F^{\mu\nu} u_\nu ~~,~~ u^\mu=\frac{dx^\mu}{ds} \,.
\label{eq:motion_in_em_field}
\end{equation}

The action for the electromagnetic field is
\begin{equation}
S_\text{em}= \int d^3x dt \ ~\frac{\sqrt{-g}}{16\pi\alpha}~F^{\mu\nu}F_{\mu\nu}
= \int d^3x dt \ ~\frac{\sqrt{-g}}{16\pi\alpha}~F_{\alpha\beta}F_{\mu\nu}g^{\alpha\mu}
g^{\beta\nu}\,,
\label{eq:em_actionE}
\end{equation}
where the dimensionless parameter $\alpha$ is the logarithmically running
coupling -- the fine structure `constant'. In effective theories,  $1/\alpha$  naturally emerges as logarithmically divergent factor, and in principle it is space- and time-dependent. For example, in quantum electrodynamics with massless fermions which emerges in superfluid  $^3$He-A \cite{Volovik2003} one has  
$1/\alpha \propto \ln [1 /  (F^{\mu\nu}F_{\mu\nu})]$.

With the choice \eqref{eq:em_dim}, the actions \eqref{eq:em_actionE} and \eqref{eq:interaction} are dimensionless:
\begin{equation}
[S_\text{int}]=[S_\text{em}]=1 \,.
\label{eq:action_dim}
\end{equation}

In effective theory,  voltage (the difference of electric potentials)  has the same dimension as frequency and electric current (the current of electrons):  $[J_e]=[A_0]=[\omega]=[t]^{-1}$. The electric resistance and conductance are dimensionless:  $[R]=1$. This suggests the possibility that the dimensionless relations between voltage and frequency,  between voltage and current, and between current and frequency  may have quantized values in some systems. The corresponding Josephson effect, quantum Hall effect, and quantum pumping, which form the so-called metrology triangle are discussed in Sec. \ref{Topological_quantum_numbers}.

\subsection{Action for gravity}

The gravitational action in effective theories is
\begin{equation}
S_\text{grav}= 
\int d^3x dt\sqrt{-g}\left(\Lambda + \frac{K}{16\pi }{\cal R}+\ldots \right) \,,
\label{eq:gravity_action}
\end{equation}
where $K$ and $\Lambda$ are gravitational coupling and cosmological constant respectively, and dots denote the higher order terms which include in particular the ${\cal R}^2$ terms.
Using dimensions of the metric component
\begin{equation}
[g_{00}] =[\hbar]^{-2} ~~,~~[g^{00}] =[\hbar]^{2}~~,
~~[g_{ik}] =[\hbar c]^{-2}~~,~~[g^{ik}] =[\hbar c]^{2}
~~,~~[\sqrt{-g}] =[\hbar]^{-4} [c]^{-3}\,,
\label{eq:g_dimension}
\end{equation}
one obtains that the dimension of the curvature  
${\cal R}$ is:
\begin{equation}
[{\cal R}] =[M]^{2}~~,~~
[d^3x dt]\,[\sqrt{-g}]
= [M]^{-4}\,.
\label{eq:R_dim}
\end{equation}
 As follows from effective theories,  dimensions
 of the gravitational coupling and cosmological constant are 
\begin{equation}
[K] =[M]^{2}~~,~~[\Lambda] =[M]^{4}\,,
\label{eq:K_dim}
\end{equation}
while the ${\cal R}^2$ terms have dimensionless prefactors.
As a result, the gravitational action is dimensionless, 
$[S_\text{grav}]=1$. The first two terms in the  gravitational action \eqref{eq:gravity_action} contain parameters with dimension $M^n$ and thus they emerge only if
conformal invariance is violated and the fundamental length or energy scale enters
the underlying microscopic theory.

The dimension of the metric suggests that metric is not the quantity,
which describes the space-time,
but the quantity, which determines the dynamics of effective fields in the background of a given quantum vacuum.

\section{Parameters of effective theory and invariant quantities}

Let us consider physical quantities which in principle can be measured, and express them in terms of the invariant parameters, which enter the action.
They can be distributed in the following groups: (i) quantities, which contain
$\hbar$ and $c$ due to historical reasons, and do not contain these parameters when expressed in terms of the parameters emerging in effective theory;
(ii) quantities, which still contain
$\hbar$ even when expressed in terms of the effective theory parameters, but do not contain it after the rescaling of the metric in \eqref{eq:g_dimension}; (iii) dimensionless geometrical and topological charges.

\subsection{quantities containing $\hbar$ due to historical reasons}

Effective theory emerging in the low energy corner contains such parameters
as fine structure constant $\alpha$, gravitational coupling  $K$, angular momentum quantum number $j$, charge quantum number $q$,  and rest energies $M$ of particles, etc.
 However, in the traditional description, which reflects the historical process of development of physical ideas, these quantities are splitted into
the electric charge of a system $Q$, elementary electric charge $e$, speed of light $c$, Newton constant $G$,
Planck constant $\hbar$, angular momentum $J$  and masses $m$:
 \begin{equation}
K= \frac{\hbar c^5}{G}~~, ~~\alpha=\frac{e^2}{\hbar c}~~,~~M=mc^2~~,~~j=\frac{J}{\hbar}~~,~~q=\frac{Q}{e}\,.
\label{eq:splitting}
\end{equation}
In effective theories such splitting is not justified, and moreover it is not necessary since the measured quantities do not contain the traditional parameters explicitly. There are some examples below.

\subsubsection{Electron energy in a Coulomb field}
\label{sec:Electron_Coulomb}

 The energy levels of electron in the Coulomb field of proton \cite{WeinbergBook}:
 \begin{equation}
M_{n,j} = M_e\left(1- \frac{\alpha^2}{ 2n^2}
-\frac {\alpha^4}{2n^4}\left(\frac{n}{ j+1/2}-\frac{3}{4}\right)+\ldots\right)\,.
\label{eq:AtomicLevels}
\end{equation}
It is expressed via  the rest energy of a free electron $M_e$;
the fine structure constant  $\alpha$; quantum number $n$
and  angular momentum quantum number $j$. 

\subsubsection{Energy levels in a Newton potential}
\label{sec:Newton_levels}

Rest energy of system of two point bodies interacting via
 Newton gravitational potential:
 \begin{equation}
 M_n= (M_1+M_2) \left(1 - \frac{1}{ 2n^2K^2} \frac{M_1^3 M_2^3}{( M_1+M_2)^2}
  +\ldots  \right)  \,.
\label{eq:GravityBoundStates}
\end{equation}
It contains rest energies of the bodies $M_1$ and $M_2$; and
gravitational coupling  $K$, which enters Einstein action
in \eqref{eq:gravity_action}.

\subsubsection{Black-hole temperature and entropy}
\label{sec:BH}

Hawking temperature of a black hole with rest
energy $M_\mathrm{BH}$
and its Bekenstein-Hawking entropy are:
 \begin{equation}
  T_\text{BH}=\frac{K}{8\pi M_\mathrm{BH}}~~,~~S_\text{BH}
             = \frac{4\pi M_\text{BH}^2}{K}   \,.
\label{eq:HawkingT}
\end{equation}

\subsubsection{Charged rotating black hole}

Entropy of the rotating electrically charges black hole:
 \begin{equation}
  S_\text{BH}(M, j, q)=\pi~\left(  2 M^2/K   -
q^2\alpha+
 2 \sqrt{M^4/K^2- j(j+1) - q^2\alpha }\right)  \,.
\label{eq:BHentropy}
\end{equation}
Here  $j$ is the angular momentum quantum number of black hole;   and $q$ is its electric charge quantum number, i.e. the dimensional charge determined by the corresponding gauge group. The charge $q$ is rational number with $q_e=-1$
for electron.

The above examples demonstrate that parameters $Q$, $e$, $c$,  $G$, $\hbar$, angular momentum 
$J$  and masses $m$ are artificial. They reflect the long history of studies of the laws of physics, but they do not appear in effective theories where all the physical laws naturally and simultaneously emerge in the low-energy corner.

\subsubsection{QCD vacuum energy}
\label{sec:QCD}

When weak or strong interaction is added, one also finds that strong and weak charges $g_S$ and $g_W$ do not enter explicitly. The Standard Model contains the corresponding running couplings and parameters of dimension $M^n$.  Example is the 
vacuum energy density -- cosmological constant -- suggested in the QCD model \cite{KlinkhamerVolovik2008d}:
\begin{equation}
 \Lambda \sim \frac{K_\text{QCD}^3}{K}
 \,.
\label{eq:InducedLambda3}
\end{equation}
It contains the gravitational coupling $K$  and  string tension  $K_\text{QCD}$ with $[K_\text{QCD}]=[\Lambda_\text{QCD}]^2=[K]=[M]^2$, where $\Lambda_\text{QCD}$ is the energy scale of QCD.

\subsection{quantities which do not contain $\hbar$ after rescaling of  metric}

\subsubsection{Zeeman energy}

 The observable consequence of quantum electrodynamics 
 is the Zeeman splitting of electron and muon energies in magnetic field $B$:
\begin{equation}
 E_\mathrm{Zeeman} =\frac{B}{M} \left( 1 +\frac{\alpha}{2\pi} +\ldots  \right)  \,,
\label{eq:Zeeman}
\end{equation}
where $M$ denotes either the electron $M_e$ or muon $M_\mu$ rest energy.
Here the field strength $F_{ik}$ is given in geometric units in \eqref{eq:em_dim}, $[F_{ik}]=[l]^{-2}$. 
The quantity $B^2-E^2=F_{\mu\nu}F^{\mu\nu}$ contains the metric elements,  
which enter $F^{\mu\nu}$. Since the dimensions of the metric elements are given by \eqref{eq:g_dimension}, the dimension of $B$ is $[B]=[M]^2$, and \eqref{eq:Zeeman} contains only the parameters  $M$ and $\alpha$.  There is one more parameter, the charge, which is not shown explicitly because 
for electron and muon one has $q_e^2=q_\mu^2=1$.

In traditional units one has  (leaving only the first term): 
\begin{equation}
 E_\mathrm{Zeeman} = \frac{e\hbar}{m} B    \,.
\label{eq:Zeeman_trad}
\end{equation}

\subsubsection{Unruh effect}
\label{sec:Unruh-effect}

In effective theory, the Unruh temperature of the  accelerated  body is
\begin{equation}
T_\mathrm{U}=\frac{ a}{2\pi}\,,
\label{eq:Unruh}
\end{equation}
where $a$ is covariant acceleration:
\begin{equation}
a^2 =
g_{\mu\nu}\frac{d^2 x^\mu}{ds^2}\frac{d^2 x^\nu}{ds^2} \,.
\label{eq:acceleration}
\end{equation}
Its dimension is $[a]=[M]$. In the traditional units, where the covariant acceleration has dimension $[a]=[l][t]^{-2}$, one has
\begin{equation}
T_\mathrm{U}=\frac{\hbar a}{2\pi c}  \,,
\label{eq:Unruh2}
\end{equation}
while in \eqref{eq:Unruh} the Planck constant $\hbar$ and $c$ do not enter explicitly, because these parameters are absorbed by the metric $g_{\mu\nu}$.

\section{Topological quantum numbers}
\label{Topological_quantum_numbers}

Topological quantum numbers in action can be considered on the example of $\theta$
term, quantum Hall effect (QHE) and Josepson effect.

\subsection{$\theta$--term}

The $\theta$ term in action
\begin{equation}
S_\theta=\frac{\theta}{16\pi^2}\int
d^3x dt
~e^{\alpha\beta\mu\nu} F_{\alpha\beta}  F_{\mu\nu}  \,.
\label{eq:theta_action}
\end{equation}
It does not contain metric and is automatically dimensionless, $[S_\theta]=1$,
 due to equations
\eqref{eq:em_dim} which
reflect the geometric nature of gauge fields.

The dimensional reduction of the $\theta$-term gives
the topological term of Chern-Simons type
in the $2+1$ action:
\begin{equation}
S_\text{CS}=\frac{k}{8\pi}\int d^2xdt
e^{\alpha\beta\gamma}A_\alpha F_{\beta\gamma} \,,
\label{eq:CS_action}
\end{equation}
where $k$ is some dimensionless fundamental number.
This action is dimensionless, $[S_\text{CS}]=1$, and does not contain the metric field.

\subsection{Quantum Hall effect}
\label{QHE}

The quantum Hall effect (QHE) in condensed matter is characterized by a similar $2+1$ action:
\begin{equation}
S_\text{QHE}=\frac{q^2\nu}{8\pi}\int d^2xdt
e^{\alpha\beta\gamma}A_\alpha F_{\beta\gamma} \,.
\label{eq:QHE_action}
\end{equation}
Here $q$ is electric charge of fermion, which is $q=q_e=-1$ for electronic system,
the systems of electrons. 
The  dimensionless  quantity $\nu$ is some fundamental number,
which characterizes the fermionic system; 
it is typically integer being related to the
fermionic  quantum topological number (Chern number)
of  the ground state of the system.
The action \eqref{eq:QHE_action} is also dimensionless, $[S_\text{QHE}]=1$.

The corresponding current of electrons ($q^2=q_e^2=1$):
\begin{equation}
j^i=\frac{\delta S}{\delta A_i}=\frac{\nu}{2\pi}e^{0ij} F_{j0}
 \,,
\label{eq:Hall_current}
\end{equation}
is transverse to electric field; this is the Hall current with
quantized Hall conductivity:
\begin{equation}
\sigma^{q}=\frac{\nu}{ 2\pi} \,.
\label{eq:Hall_conductivity_q}
\end{equation}
In effective theory, the Hall (and also the spin-Hall) conductance are expressed in terms of integer or rational number and $\pi$.

In traditional units, where $A_0$ is substituted by $e\tilde A_0/\hbar$, and action is multiplied by $\hbar$, the Hall conductivity
is given by \cite{FlowersPetley2008}
\begin{equation}
\sigma _{xy}=4\pi \alpha\sigma^{q}= 2\alpha \nu \,,
\label{eq:Hall_conductivity}
\end{equation}
where $\alpha$ is the fine-structure constant.
In this units, $\sigma _{xy}$ is not expressed in terms of  rational number  and $\pi$. The quantization is not exact, since the fine-structure constant is not a constant but is a running coupling. It depends on the infrared cut-off and thus is space- and time-dependent.
The failure of the traditional description to obtain exact quantization comes from the unjustified splitting of the vector potential, $A_\mu = (e/\hbar) \tilde A_\mu$, in the traditional description.
As a result the field $\tilde F_{\mu\nu} = \nabla_\mu \tilde A_\nu-\nabla_\nu \tilde A_\mu$ is not gauge invariant: under gauge transformation
$\tilde A_\mu \rightarrow \tilde A_\mu + (\hbar/e)\nabla_\mu \phi$, the field transforms as $\tilde F_{\mu\nu} \rightarrow \tilde F_{\mu\nu} + \nabla_\mu
\left( \hbar/e\right)\nabla_\nu \phi  -  \nabla_\nu\left( \hbar/e\right)\nabla_\mu \phi $. The field $\tilde F_{\mu\nu}$ would be gauge invariant, only if the electric charge $e$ and $\hbar$ are the fundamental constants. But the charge $e$ is certainly not a fundamental constant. This is the so-called `physical charge', which is obtained by splitting of $\alpha$
and thus is coordinate dependent. The gauge invariance and the true quantization of Hall conductivity are restored when the geometric vector potential $A_\mu$ and the geometric $U(1)$ charge of electron  $q_e=-1$ are used.   

\subsection{Josephson effect}

\subsubsection{Josephson effect in superconductors}

The ac Josephson effect in superconductors (charged superfluids)
comes from the coupling of the phase $\phi$ of the order parameter with electromagnetic field: due to gauge invariance the time derivative of $\phi$ enters in action only in combination with the electric potential, $\partial_t  \phi-qA_0$, where $q=2q_e=-2$ is the electric charge of Cooper pairs.
Taking into account the $2\pi$ periodicity of the phase $\phi$ one obtains the following Josephson relation in superconductors:
\begin{equation}
\omega= 2\left(A_0^{(2)}-A_0^{(1)}\right) \,.
\label{eq:Josephson_e}
\end{equation}
In effective theory, the Josephson relation contains only integer $\vert q\vert =2$. This is because electromagntic field is given in geometric presentation, in which  $A_0$ has the same dimension as frequency: $[A_0]=[\omega]=[t]^{-1}$.
That is why the Josephson effect provides the standard of voltage in inverse unit of time.

In traditional units, the Josephson relation contains parameters $e$ and $\hbar$. The electric charge $e$ is certainly non-fundamental, while $\hbar$
can be non-fundamental. In both cases the quantization is lost, as in the case of quantum Hall effect in Sec. \ref{QHE}.

\subsubsection{Josephson effect in neutral systems}

Let us now consider  the ac Josephson effect 
in electrically neutral systems (superfluids).
In superfluid liquids, such as superfluid  $^4$He and $^3$He, the role of voltage can be played by static gravitational field.  That is why let us take into
account gravity. In general relativity there is the Tolman law for temperature and for the chemical potential in thermodynamic equilibrium in the presence of a static gravitational field. It relates the local values $T({\bf r})$, $\mu({\bf r})$ with the global values $T_\mathrm{T}$ and $\mu_\mathrm{T}$ which are space-independent in equilibtium:
\begin{equation}
T({\bf r})\sqrt{-g_{00}({\bf r}) }=T_\mathrm{T} ~~,~~ \mu({\bf r})\sqrt{-g_{00}({\bf r}) }=\mu_\mathrm{T} ~,
\label{eq:MassTolmanLaw}
\end{equation}
The dimensions of the local and global quantities are
\begin{equation}
[T]=[ \mu]=[M] ~~,~~[T_\mathrm{T}]=[\mu_\mathrm{T}]=[\omega]= [t]^{-1}\,.
\label{eq:Josephson_dim}
\end{equation}
In neutral superfluids the Josephson oscillations emerge due to difference of chemical potentials. Oscillations cannot emerge in equilibrium:
they only occur if the Tolman potential have a jump
across the Josephson junction.
That is why the correct Josephson relation in neutral superfluids should be:
\begin{equation}
\omega= \nu\left(\mu_\mathrm{T}^{(2)} -\mu_\mathrm{T}^{(1)}\right)  \,,
\label{eq:JosephsonTolman}
\end{equation}
where integer $\nu=1$ for superfluid  $^4$He  and $\nu=2$ for superfluid  $^3$He with Cooper pairing mechanism of superfluidity. In traditional units,
the parameter $\hbar$ enters explicitly, and if $\hbar$ is not fundamental, the quantization is not exact. This in principle will allow us to experimentally resolve between fundamental and non-fundamental $\hbar$.

In terms of the local chemical potential $\mu$, equation
\eqref{eq:JosephsonTolman} becomes (we use $\nu=1$ assuming  superfluid  $^4$He):
\begin{equation}
\omega= \sqrt{-g_{00}^{(2)} } \mu^{(2)}-\sqrt{-g_{00}^{(1)} } \mu^{(1)} \,.
\label{eq:Josephson_mu}
\end{equation}
In case when gravity is the same across the junction, and only the values of the local chemical potential $\mu$ 
are different on two sides of the contact, the Josephson relation
\eqref{eq:Josephson_mu} becomes
\begin{equation}
\omega= \sqrt{-g_{00} }( \mu^{(2)}- \mu^{(1)}) \,.
\label{eq:Josephson_mu2}
\end{equation}
In Minkowski vacuum this equation obtains the  traditional form $\omega= ( \mu^{(2)}- \mu^{(1)})/\hbar$. 

In the other case, when the chemical potential is the same on both sides
of the contact, but the gravitational potential is different, the equation \eqref{eq:Josephson_mu}  
becomes
\begin{equation}
\omega=\mu \left( \sqrt{-g_{00}^{(2)} }- \sqrt{-g_{00}^{(1)} } \right)\,.
\label{eq:Josephson_mu3}
\end{equation}
This situation takes place when the Josephson junction is represented by a channel connecting two vessels in which the level of the liquid 
has different height, $h_2$ and $h_1$. In the limit of a weak gravitational potential in the background of Minkowski space-time, one has $g_{00}\approx -(1-2\Phi/c^2)/\hbar^{2}$, $ \sqrt{-g_{00}^{(2)} }- \sqrt{-g_{00}^{(1)}} \approx 
-(\Phi ^{(2)} -\Phi ^{(1)})/\hbar c^2$, $\Phi ^{(2)} -\Phi ^{(1)}=g(h_2-h_1)$, and the 
equation \eqref{eq:Josephson_mu3} transforms to the the familiar exression for the Josepson effect in superfluid $^4$He caused by gravitational field:
\begin{equation}
\omega=\frac{mgh}{\hbar}~~,~~h=|h_2-h_1|\,.
\label{eq:Josephson_mu4}
\end{equation}

In superfluid $^4$He in the limit of zero temperature, the chemical potential 
equal the rest energy per unit atom of $^4$He: $\mu=M$. This $M$ slightly differs from the rest energy of an isolated $^4$He atom due to the energy added by interaction between the atoms and zero point motion of $^4$He atoms in superfluid, $M\neq M_4$.  That is why in effective theory, the Josephson relation \eqref{eq:Josephson_mu2} has the same form as both the gravitational red shift and the energy-frequency relation in quantum mechanics
\eqref{eq:Emission_covariant}:
\begin{equation}
\omega({\bf r})= \sqrt{-g_{00}({\bf r})} \left(M^\mathrm{i}- M^\mathrm{f} \right)\,,
\label{eq:red_shift}
\end{equation}
where $M^\mathrm{i}$ and $M^\mathrm{f}$ are the rest energies of an atom
in initial state before radiation of photon and in the final state
after the radiation correspondingly. 

\subsection{Metrology triangle}

Another topological effect is quantum pumping, the transfer of fermions by periodic
change of the parameters of the system: $\dot N=\nu \omega/(2\pi)$, where  $\dot N$ is the number of fermions transferred per unit time between two subsystems, and $\nu$ is topological quantum number.  The
quantum pumping in electronic systems (single-electron tunnelling  \cite{FlowersPetley2008,Pekola2008}) reflects the quantization of the number of electrons. Since electric current is the charge $q$ transferred per unit time, $J=q_e\dot N$, one obtains the relation between the current and frequency:    $J=q_e\nu \omega/(2\pi)$. The quantum pumping allows to calibrate frequency by measuring the current $J$, or to calibrate current by tuning frequency.  This completes the so-called metrology triangle:
Josephson effect, QHE and quantum pumping relate repsectively voltage and frequency, current and voltage, and frequency and current.

In effective theory, the current $J$, the voltage $\Delta A_0$ and frequency $\omega$ are all expressed in inverse time units. The relations between these quantities are determined solely by the fundamental geometric or topological charges, integer or fractional, such as $q$ and $\nu$. These three effects reflect different geometrical and topological properties of condensed matter system and provide two independent ways of calibration of voltage in terms of frequency: either directly by Josepson effect or by combination of QHE and quantum pumping. 

Using the standard of voltage one may measure the potential produced by the geometric charge $q$: 
\begin{equation}
A_0=\frac{\alpha q}{r} \frac{1}{g^{rr}g^{00}\sqrt{-g}}
 \,.
\label{eq:PotentialOfCharge}
\end{equation}
This allows to measure the combination of the metric elements after $\alpha$
is measured using \eqref{eq:AtomicLevels}.
The other experiments measure other parameters and other combinations of the metric tensor. For example, $g_{00}$ can be found using equation \eqref{eq:Emission}.

\section{Schr\"odinger equation}

\subsection{Klein-Gordon equation for scalar  field}

In effective theory, the dimensionless action for a scalar field $\Phi(x)$ with dimension  $[\Phi]=[M]$
has the form:
\begin{equation}
S_\text{scalar}=\frac{1}{2}\int d^3x dt \ \sqrt{-g}\;
\left(g^{\mu\nu}\nabla_\mu\Phi^*\,\nabla_\nu \Phi -  M^2|\Phi|^2\right) \,.
\label{eq:Klein_Gordon}
\end{equation}
The kinetic term has the same dimension as the mass term without
artificial introduction of the parameter $\hbar$.

\subsection{Schr\"odinger equation}

The nonrelativistic Schr\"odinger action can be obtained by expansion of Eq. \eqref{eq:Klein_Gordon}. In case of space-time independent $g^{00}$ one introduces the Schr\"odinger wave function $\Psi$
 with dimension  $[\Psi]=[M]^{3/2}$
\begin{equation}
\Phi({\bf r},t) = \frac{1}{\sqrt{M}}\exp\left(i Mt /\sqrt{-g^{00}}\right)\Psi({\bf r},t)  \,.
\label{eq:PhiPsi}
\end{equation}
After expansion over $1/M$ one obtains the Schr\"odinger-type  action in the form
\begin{eqnarray}
S_\text{Sch}=\int d^3x dt  \sqrt{-g} L~,
\nonumber
\\
2L= 
i\sqrt{-g^{00}} \left(\Psi \partial_t \Psi^*-\Psi^* \partial_t \Psi\right)
+  \frac{ig^{0k}}{\sqrt{-g^{00}} }\left(\Psi \nabla_k \Psi^*-\Psi^* \nabla_k \Psi\right)
+\frac{g^{ik}}{M}\nabla_i\Psi^* \nabla_k \Psi   \,.
\label{eq:Schroedinger}
\end{eqnarray}

For Minkowski space-time, \eqref{eq:Schroedinger} transforms after renormalization $\Psi \rightarrow \Psi (\hbar c)^{3/2}$
 to
\begin{equation}
S_\text{Sch}^\text{Mink}=\frac{1}{2}\int d^3x dt
\left(
i \left(\Psi \partial_t \Psi^*-\Psi^* \partial_t \Psi\right)
+\frac{\hbar}{m}\nabla_i\Psi^* \nabla_i \Psi
    \right)  \,.
\label{eq:Schroedinger_trad}
\end{equation}
The equation 
 \eqref{eq:Schroedinger_trad} contains a single parameter: the ratio of parameters  $\hbar$ and $m$.
 In non-relativistic quantum mechanics, they always enter together. Examples are equation \eqref{eq:Josephson_mu4} for the Josephson effect in neutral superfluids and also the quantum of circulation of superfluid velocity: $\kappa = 2\pi \nu \hbar/m$. Here $m$ is the mass of an atom in a superfluid liquid; $\nu=1$ for superfluid $^4$He;  $\nu=1/2$ for superfluid $^3$He-B; and $\nu=1/4$ for the four-particle correlated state, which may occur in thin superfluid films, see e.g. Section 10 in Ref. \cite{Volovik1992}.
 
  The  traditional expression for non-relativistic Schr\"odinger action  is obtained after multiplication of  the  action \eqref{eq:Schroedinger_trad}  by $\hbar$. It contains the parameters of the equilibrium  Minkowski vacuum,    $c$ (via $m=M/c^2$) and  $\hbar$. In the traditional form
  the spectrum of a nonrelativistic particle with mass $m$ in the 1D box of size $L_x$ with inpenetrable walls is 
  \begin{equation}
E_n=\frac{\hbar^2\pi^2n^2}{2mL_x^2}\,.
\label{eq:spectrum_in_box_trad}
\end{equation}
However, in the covariant form it is
\begin{equation}
E_n=\frac{\pi^2n^2}{2M(\Delta l)^2}\,,
\label{eq:spectrum_in_box}
\end{equation}
where $\Delta l$ is the proper distance, $(\Delta l)^2=L_x^2g_{xx}$. Being written in the covariant form,
the energy spectrum does not contain parameters $m$, $c$ and $\hbar$ explicitly. This demonstrates that
even the non-relativistic quantum mechanics contains the rest energy $M$ instead of traditional
mass $m$. This suggests that the Lorentz invariance and general covariance are not very important
factors. The metric field $g_{\mu\nu}$ may serve the source of parameters $\hbar$ and $c$ even for
the so called one half of general relativity \cite{Visser0309072}, when
the matter fields feel the metric field as effective geometry, but $g_{\mu\nu}$
does not necessarily obey the Einstein equations.

\section{Discussion. Emergent vs fundamental physics}

In general relativity, the parameters $\hbar$ and $c$ vanish from equations written in covariant form. This statement can be trivial  or not, depending on whether gravity
and other physical laws are fundamental or emergent.

In fundamental physics, the parameter $\hbar$ is a universal constant,
that is why, in principle it
can be hidden in the metric.  It is important that,
after it is included into the metric,
$\hbar$ completely disappears from all physical equations if the action is represented in terms of the natural parameters.
This could mean that $\hbar$  is the natural part
of the metric, just in the same manner as the parameter $c$, which 
 is artificially introduced for convention together with
Boltzmann constant $k_B$ 
\footnote{Let us recall, that in fundamental general relativity there is no limiting velocity. For example, in expanding Universe the coordinate velocity of galaxies may exceed the speed of light, due to the `aether drift'.  Only the relative velocity of two objects  in the same point of space-time is limited by $c$. But  GR
does not discriminate between Minkowski metrics with different  $c$. On the contrary, in effective GR the existence of distinguished Minkowski metric with fixed parameters is typical.}
The Planck constant $\hbar$ becomes the second parameter  of
a particular two-parametric family of solutions of
general relativity equations describing the flat isotropic Minkowski vacuum
in \eqref{eq:Minkowski_metric}.
After $\hbar$ becomes the part of the metric, it shares the
same fate as parameter $c$.
Since GR  does not discriminate between metrics with different
$\hbar$, the parameter $\hbar$  becomes the matter of convention too.
Thus if GR is fundamental, the inclusion of $\hbar$ into the metric  is equivalent to the choice of units  $\hbar=c=1$, which is trivial and does not lead to any new results.   

In the emergent theories, there are no universal constants: the  `fundamental' constants may be given by vacuum expectation values of the $q$-field \cite{KlinkhamerVolovik2008b},  of scalar fields 
 \cite{Trialogue}, of a 4-index field strength \cite{Trialogue,KlinkhamerVolovik2008b}, etc.
That is why $\hbar$ and $c$ could and should be space/time dependent 
in the same manner as $\alpha$.
At first glance, the space-time dependence of the `fundamental constant'  $\hbar$ would violate many physical laws. This happens if the equations are written in the traditional from. The traditional form of the Feynman integral over the fields $\chi$ is
$\int d\chi \exp\left( iS/\hbar\right)$.  The quantum-mechanical phase
factor is a compact U(1) quantity, which means that action acquires topological properties.
If  $\hbar$ is space-time dependent, it must be
introduced within the integral:
$S/\hbar \rightarrow \int d^3x dt~ \hbar^{-1}\sqrt{-g}L$.
But this would violate the general covariance. This violation does not occur if
$\hbar$ is a scalar field, but even in this case  the gauge invariance of the topological terms in the action would be violated (see Sec. \ref{Topological_quantum_numbers}).
That is why in the emergent  gravity, it is simply necessary
to include $\hbar$ into the metric in order to avoid such violations.
As a result the action becomes dimensionless,
and the  Feynman integral
becomes
$\int d\chi \exp\left( i S\right)$. Just in the same way the  gauge invariance requires the use of the geometric electric charge $q$ instead of the traditional electric charge $e$, which was introduced at the earlier stage of physics. The latter must be included into the geometric vector potential $A_\mu$ and into the fine structure `constant' $\alpha$.

In the emergent theories,  the quantities $\hbar$ and $c$ do not enter the equations of general relativity, but they are two parameters which characterize the special solution of the equations. This solution describes the effective Minkowski metric emerging in the background of the equilibrium vacuum state of our Universe. In case if the equilibrium vacuum state corresponds to the effective de Sitter space-time,  one more parameter will characterize our vacuum, the Hubble constant $H$ which enters the effective interval in the de Sitter Universe:
\begin{equation}
ds^2=\frac{1}{\hbar^2}\left(-dt^2+ \frac{1}{c^2}(d{\bf r}-H{\bf r}dt)^2 \right)  \,.
\label{eq:de_Sitter}
\end{equation}
The parameter $H$ also does not enter the GR equations, and in this respect it is similar to $c$ and 
$\hbar$. However, as distinct from Minkowski vacuum, de Sitter vacuum can be dynamically unstable
\cite{KlinkhamerVolovik2009c} and thus its Hubble parameter cannot serve as fundamental constant. 

In conclusion, the fundamental gravity is equivalent to the universal $\hbar$,
while in
emergent gravity  $\hbar$ is non-universal.
In the former case $\hbar$ \underline{can} be included into metric, while
in the latter case $\hbar$ \underline{must} be included into metric.
Inclusion of $\hbar$ into metric suggests
that $\hbar$ becomes related to gravity (maybe to dilaton field
in effective gravity)
and this
reflects the peaceful coexistence and
interplay of gravity and quantum mechanics.

Finally, let us mention an example, where the difference between the fundamental and effective theories
is most pronounced. If Minkowski vacuum emerges as a result of spontaneous symmetry breaking, it can be degenerate with different signs of emergent parameters $\hbar$ and  $c$. The bosonic fields cannot distinguish between the vacua with different signs of $c$ and $\hbar$, because these quantities enter quadratically into the metric. For fermions these 4 vacua are physically different, because their dynamics is determined by the vierbein, rather than by metric, and vierbein depends linearly on $\hbar$ and $c$. The physical laws, however, are the same in all these vacua, if the underlying symmetry which connects these vacua is exact, and in all the equations discussed here $\hbar$ must be substituted by $\sqrt{\hbar ^2}$. The interesting consequence of such symmetry breaking would be the possibility of domain walls separating the vacua with different signs of $\hbar$ and  $c$: the  $\hbar$-wall where $\hbar$ changes sign; the corresponding  $c$-walls; and combined walls. Within the  $\hbar$-wall, the parameter $\hbar$ crosses either the value $\hbar=0$ or the value $\hbar=\infty$. In condensed matter the analogs of such walls in the vierben field can be found in Ref. \cite{Volovik1990}; these include  the walls in which one, two or all the three speeds of light in Eq. \eqref{eq:Anisotropic_Minkowski} change sign (see also Ref. \cite{Volovik2009} and references therein).  

It is a pleasure to thank Frans Klinkhamer, Gordey Lesovik and Yuriy Makhlin for discussions and criticism, and Michael Duff for e-mail correspondence. This work is supported in part by the Russian Foundation
for Basic Research (grant 06--02--16002--a) and the
Khalatnikov--Starobinsky leading scientific school (grant
4899.2008.2).

\end{document}